\documentclass[preprint,review,12pt]{elsarticle}

\usepackage{graphicx}
\usepackage{amssymb}
\usepackage{multirow}
\usepackage{array}
\usepackage{xcolor}
\usepackage{caption}
\usepackage{subcaption}
\usepackage{graphicx}
\usepackage{lineno}

\journal{Advances in Space Research}

\begin{document}

\begin{frontmatter}


\title{Systematic Study of Ionospheric Scintillation over the Indian Low-Latitudes during Low Solar Activity conditions}



\author[a]{Deepthi Ayyagari\corref{c-d54cc1eb1ca4}}
\ead{nagavijayadeepthi@gmail.com}\cortext[c-d54cc1eb1ca4]{Corresponding author.}

\author[a]{Abhirup Datta}
\ead{abhirup.datta@iiti.ac.in}

\author[a,b]{Sumanjit Chakraborty}
\ead{sumanjit11@gmail.com}

\address[a]{Department of Astronomy, Astrophysics and Space Engineering\unskip, Indian Institute of Technology Indore\unskip, Simrol \unskip, Indore\unskip, 453552, Madhya Pradesh, India}

\address[b]{Space and Atmospheric Sciences Division\unskip, Physical Research Laboratory\unskip, Ahmedabad 380009\unskip, Gujarat, India.}

\begin{abstract}

A systematic study of ionospheric scintillation at the low-latitudes, especially around the Equatorial Ionization Anomaly (EIA) and the magnetic equator, is essential in understanding the dynamics of ionospheric variation and related physical processes. Our study involves NavIC $S_{4_C}$ observations over Indore and Hyderabad. Additionally, GPS $S_{4_C}$ observations over Indore were analyzed, under disturbed as well as quiet time ionospheric conditions from September 2017 through 2019, falling in the declining phase of the solar cycle 24. The $S_{4_C}$ observations were further analyzed using proxy parameters: ROT and ROTI. These results have been obtained from three satellites of the NavIC constellation (PRNs 2, 5, and 6). The onset times of scintillations \textbf{were} observed to be around 19:30 LT (h) and 20:30 LT (h) for Hyderabad and Indore respectively, while the $S_{4_C}$ peak values occurred between 22:00 LT (h) and 23:00 LT (h). The reliability of NavIC was evaluated using scattering coefficients that revealed a good correlation across the pair of signals during quiet time ionospheric conditions. The observations clearly show that the amplitude scintillation of the NavIC signal follows the Nakagami-m distribution along with the $\alpha-\mu$ distribution as a depiction of the deep power fades caused by scintillation on these signals. This paper shows the impact of such systematic studies near these locations for the first time, in improving the understanding of the dynamic nature of low-latitude ionosphere under low solar activity conditions.

\end{abstract}

\end{frontmatter}


\section{Introduction}
\label{S:1}

The Earth's ionosphere, which is approximately stratified between 60 and 1000 km above ground, often produces electron density disturbances, which disperse radio waves passing through it, resulting in amplitude and phase scintillations. The degree to which the ionospheric irregularities cause scintillations is determined by the frequency of the signal and the intensity of the electron density irregularities. In general, ionospheric scintillation is described as a rapid or abrupt change in phase and amplitude of a radio signal, while traversing through the ionosphere \citep{21,16,17,18,19,20}. This phenomenon affects radio wave signals with frequencies ranging from 100 MHz to 4 GHz in general \citep{2}. Several applications are affected by the ionospheric scintillations including positioning, navigation, communication, and remote sensing \citep{1,3,5}. Owing to the variability of factors such as solar activity, interplanetary magnetic field conditions, local electric field, conductivity, convection processes, and wave interactions, ionospheric scintillation is difficult to predict and model in the equatorial and low-latitude regions, characterized by the Equatorial Ionization Anomaly (EIA) \citep{9,6,7,8}. Ionospheric scintillations are typically associated with high solar activity and geomagnetic storms at high latitudes, while disruptions in and around the doldrums are primarily caused by the EIA and ionospheric bubbles or Equatorial Plasma Bubbles (EPBs), which cause extreme scintillation in radio wave communications from satellites to receiver location on ground \citep{10,11,22,61,67}.

Several researchers have studied the phenomenon of ionospheric scintillation and its causes and effects during extreme geomagnetic activity over the past few decades using Global Positioning System (GPS) data. \citep{19} was one of the first studies to show, how the number of sunspots and geomagnetic latitude affect radio star and satellite scintillations. The morphology of this global phenomenon is clearly explained by \citep{4}. It is possible to characterize and comprehend the ionospheric disturbances that cause ionospheric scintillations using the GPS data \citep{10,11,14,15}. During the magnetic storms of September 1999 and July 2000, studies by \citep{23,27} recorded depletion of ionization density structures from middle and equatorial latitudes across the United States, where plumes of greatly enhanced Total Electron Content (TEC) were correlated with the erosion of the outer plasmasphere by turbulent sub-auroral penetration electric fields. \citep{28} was the first to report the effect of $C/N_0$ fluctuations on GPS Dilution Of Precision (DoP) for minimum sunspot number conditions under geomagnetically quiet conditions. \citep{25} reported on the occurrence of GPS cycle slips observed in a duration of exceptionally low solar activity on October 8, 2009, correlated with the phenomenon of phase scintillation, in a clear study from the Indian longitude sector. During moderate to high solar activity levels in September 2011, a simple approach to the causes of cycle slips of signals from satellite-based navigation systems from two stations on and beyond the anomaly crest region was presented by \citep{29}.

To understand the variabilities of the upper atmosphere and space weather impacts on it over the geosensitive Indian longitude sector, the Indian Space Research Organization (ISRO) conceived the regional navigation satellite system: Navigation with Indian Constellation (NavIC). The space segment of this system consists of seven satellites (a combination of Geostationary Earth Orbit (GEO) and Geosynchronous Orbit (GSO) satellites). These satellites transmit signals in the $L_5$ and $S_1$ bands, with carrier frequencies of 1176.45 MHz and 2492.028 MHz, respectively, in a 24 MHz bandwidth. It is engineered to provide positional accuracy information not only to the Indian users but also to the users within a 1500 km radius around its boundary, defined by a rectangular grid spanning from 30\ensuremath{^{o}}S to 50\ensuremath{^{o}}N in latitude to 30\ensuremath{^{o}}E to 130\ensuremath{^{o}}E in longitude \citep{36}. The reliability of NavIC in exploring the upper atmosphere has already been demonstrated by \citep{31,37,38,39,33,32,35,34,69,70,71,72,73,74}. The signal intensity and accuracy of NavIC satellites have recently been confirmed by \citep{30}, to be reliable and used for ionospheric scintillation studies. The study of observed ionospheric scintillation events from GPS and NavIC measurements in the Bengaluru region was reported by \citep{12}. Later, \citep{13} reported two-station observations of ionospheric scintillations of NavIC $L_5$ and $S_1$-band signals over the Indian subcontinent's low-latitude regions.  

The present work, to the best of our knowledge, is the first long-term study of low-latitude ionospheric scintillation, during low solar activity conditions, using observations from both NavIC and GPS signals over a wide area: Indore located near the northern crest of the EIA and Hyderabad located in between the crest and the magnetic equator. This in-depth study would set a path forward for a thorough understanding of the dynamic nature of the low-latitude ionosphere during low solar activity conditions over the Indian longitude sector. The approach for NavIC data for various other locations to validate with multiple regions study was initiated during this work. But due to the unavailability of the data which matches the analysis period and the limited network of NavIC Rx, the analysis presented is limited to these stations only.

\section{NavIC and GPS Receiver - Data Analysis}

\begin{figure}
\centering
\includegraphics[width=2.5in,height=3in]{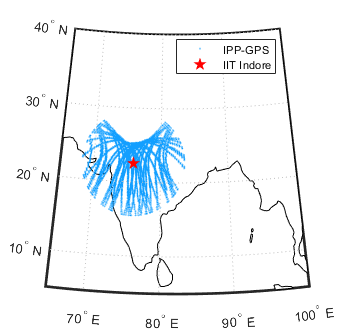}
\includegraphics[width=2.5in,height=3in]{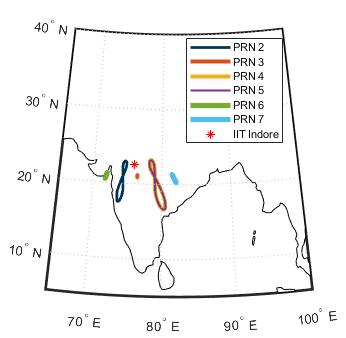}
\includegraphics[width=2.5in,height=3in]{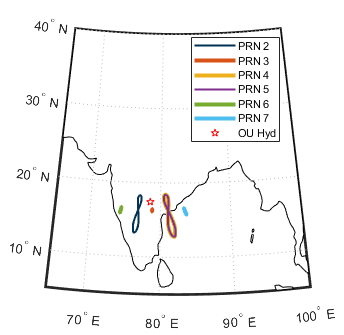}\\
\caption{The figure represents the locations of observation that are presented in the work. The red stars in each of the maps represent the receiver's location. The eight shaped plots in each of the maps are the ionospheric pierce point locations as observed from the GPS and NavIC receiver locations for Indore (top) and NavIC receiver location for Hyderabad (bottom).}
\label{fig:Fig1}
\end{figure}

In the Department of Astronomy, Astrophysics and Space Engineering (DAASE) of the Indian Institute of Technology, Indore, a multi-constellation, multi-frequency GNSS receiver (Septentrio PolaRxS Pro), capable to receive GPS $L_1$ (1575.42 MHz), $L_2$ (1227.60 MHz), and $L_5$ (1176.45 MHz) signals, as well as a NavIC receiver (ACCORD) provided by Space Applications Centre, ISRO, intercepting GPS $L_1$, NavIC $L_5$, and NavIC $S_1$ (2492.028 MHz) signals, are operational since May 2016 and May 2017, respectively. In addition to the Indore (Lat: 22.52\ensuremath{^{o}}N, Lon: 75.92\ensuremath{^{o}}E; Magnetic dip: 32.23\ensuremath{^{o}}N) station's NavIC and GPS data, the present study also involves the utilization of NavIC data from the station: Osmania University (OU), Hyderabad (Lat: 17.40\ensuremath{^{o}}N, Lon: 78.51\ensuremath{^{o}}E; Magnetic dip: 21.69\ensuremath{^{o}}N) as depicted in Fig. \ref{fig:Fig1} which shows the geographic locations of the GPS and the NavIC receivers along with the Ionospheric pierce points (IPPs) as observed from each of the locations.

The NavIC receiver generates raw log files that are post-processed to retrieve the observables. The observables are only retrieved with the aid of the Graphic User Interface (GUI) application (see for details in ACCORD-Original Equipment Manufacturer (OEM) \citep{36}). The GUI application receives, down-converts and demodulates the transmitted satellite signals both at $L_5$ and $S_1$ band frequencies. Most importantly, this GUI application generates measurements precisely with respect to the external/internal trigger, such as 1 Pulse Per Second (PPS). In addition, the GUI application includes a GPS receiver capable of receiving and processing $L_1$ Coarse/Acquisition (C/A) code, centered at 1575.42 MHz signal, and generates measurements with respect to the external/internal trigger. The receiver output provides the user position by computing only $L_5$ , only $S_1$ , only $L_1$ , $L_5$ and $S_1$ (Dual frequency) modes, and NavIC + GPS (Hybrid) mode depending on the user command. The receiver uses Satellite-based Augmented System (SBAS) corrections with the GPS automatically if SBAS correction parameters are available. Control inputs to the system include a 10 MHz external reference clock, 1-PPS signal; two RS232 interfaces for timing messages and National Marine Electronics Association (NMEA) message format, and an Ethernet port for monitoring, controlling, and commanding. The same Ethernet port is also used to do the Software Upgrade (SWUG) of the receiver. The data update rate on the NMEA port is 1 Hz and that of the Ethernet port is 5 Hz for GPS and 1Hz for NavIC. The data obtained at this rate is logged in the format of comma-separated variable (.csv) files for every three hours, using which an automated pipeline has been developed which uses observables such as: Time (TOW (s)), $C/N_0$, Azimuth, Elevation, Iono Delay, TEC and loss of lock at the sampling frequency of 1 Hz to match the ACCORD generated NavIC observables with Septentrio-generated GPS observables to generate the various plots of IPP footprints, $S_I$, $S_{4_C}$, ROT and ROTI, as well as the $S_{coeff}$ values and the various distribution curves, plots, and values.

The Septentrio PolaRxS Pro is a state-of-the-art GNSS receiver that widens the capabilities of studying the upper atmosphere by using a modern triple-frequency receiver. The tracking engine is coupled with an ultra-low noise OCXO (details given in POLARXS-OEM) frequency that generates and stores raw high rate data which can be processed in real-time or in post-processing to furnish 60s scintillation indices $S_4$ and $\sigma_\phi$, along with other parameters like TEC, lock-time and the scintillation spectral parameters such as the spectral slope and the spectral strength of the phase Power Spectral Density (PSD) at 1\,Hz for all visible satellites and frequencies. The data from this receiver usually allows the user to utilize the data in the following formats:
\begin{itemize}
    \item Highly compact and detailed Septentrio Binary Format (SBF) output.
    \item NMEA v2.30 and v4.10 output format.
    \item ISMR (Ionospheric Scintillation Monitoring) file generation using the provided sbf2ismr utility.
\end{itemize}
In the present study, Data have been utilized during the whole period (25 months) of analysis, beginning from September 01, 2017, till September 30, 2019. Data are obtained in the .ismr format files generated every hour. A more elaborate GPS data analysis procedure has been discussed in \citep{60,38}.


\section{Observations of Scintillations}

\subsection{Criteria for scintillation}

\begin{figure}
\centering
\includegraphics[width=\textwidth,height=4in]{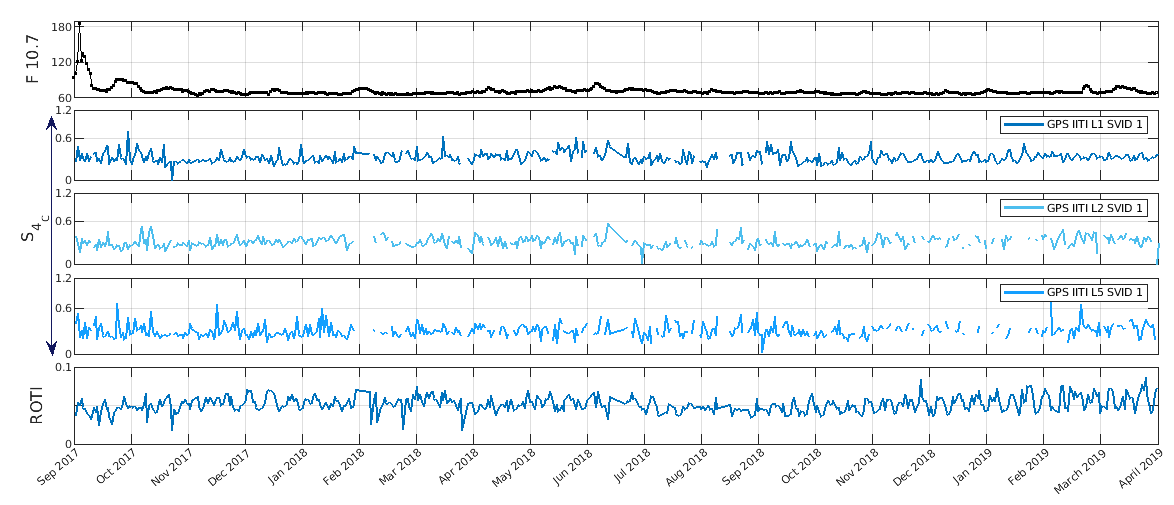}
\caption{The variation of F10.7 along with GPS SVID 1 observed values of peak $S_{4_C}$ and ROTI by SVID 1 from Indore station for the available L band frequencies, for the analysis period: September 2017 through April 2019.}
\label{fig:Fig2}
\end{figure}

\begin{figure}
\centering
\includegraphics[width=\textwidth,height=2.2in]{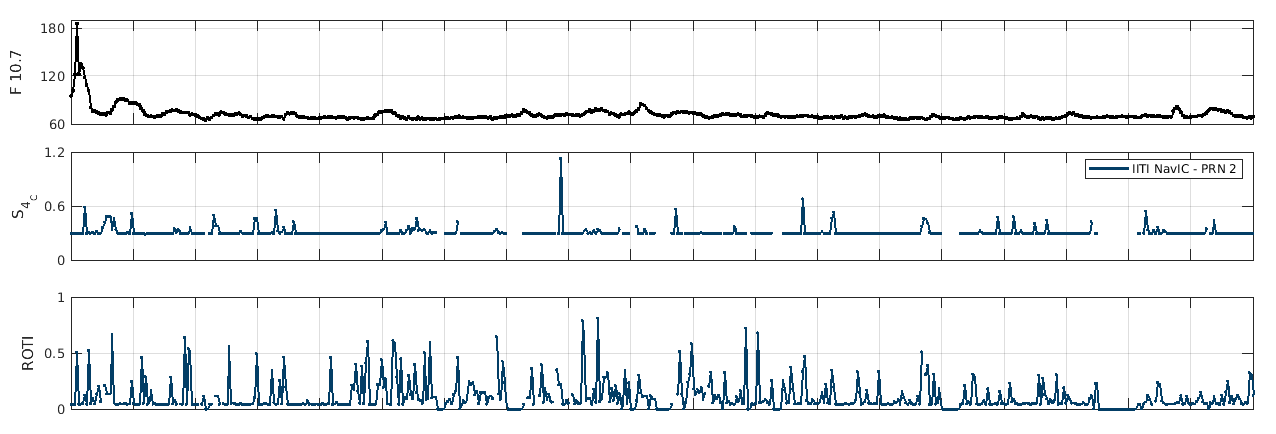}
\includegraphics[width=\textwidth,height=2.2in]{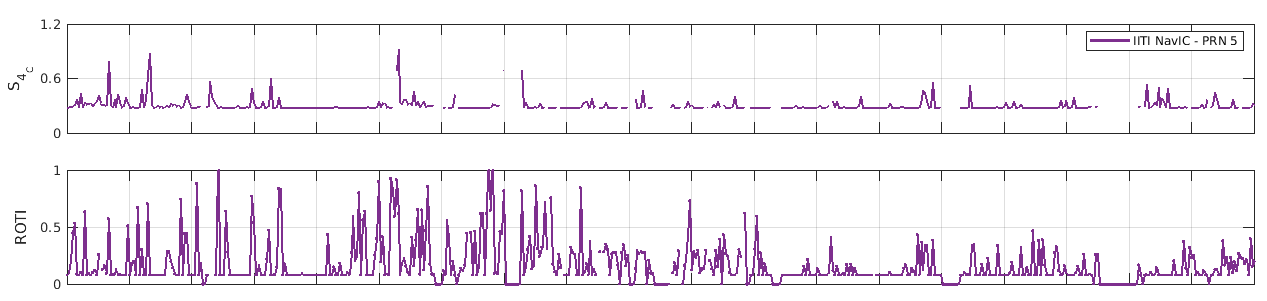}
\includegraphics[width=\textwidth,height=2.2in]{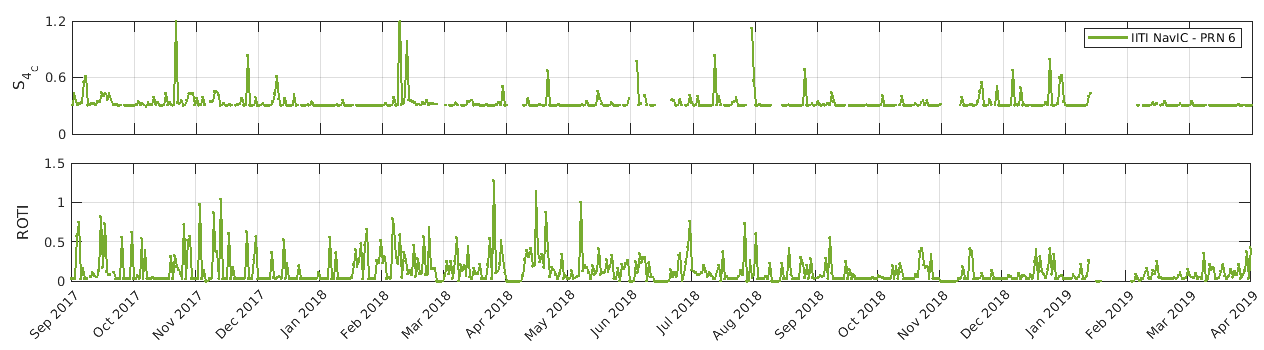}
\caption{The variation of F10.7 along with NavIC observed values of peak $S_{4_C}$ and ROTI by PRNs 2, 5 and 6 from Indore station, for the analysis period: September 2017 through April 2019.}
\label{fig:Fig3}
\end{figure}

\begin{figure}
\centering
\includegraphics[width=0.95\textwidth,height=2.2in]{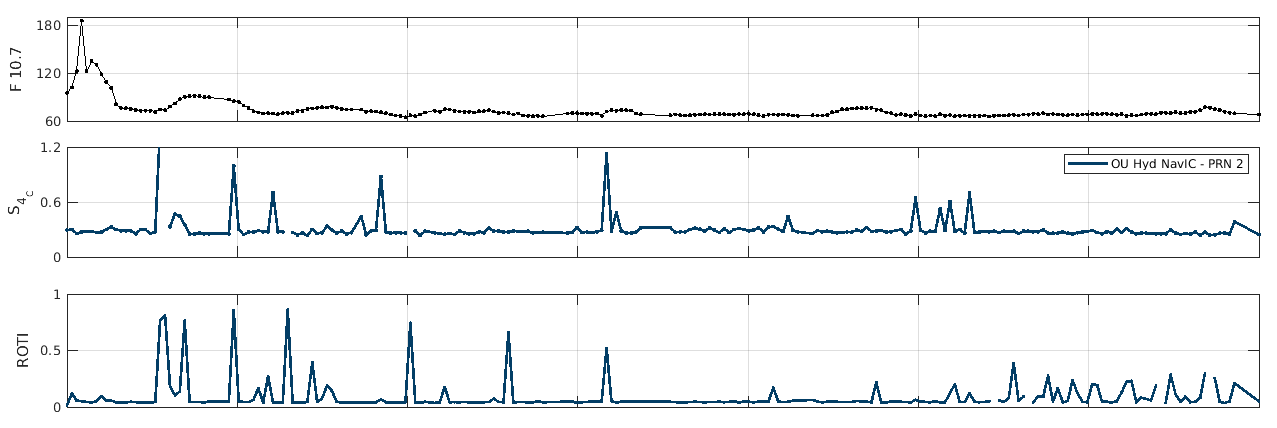}
\includegraphics[width=0.95\textwidth,height=2.2in]{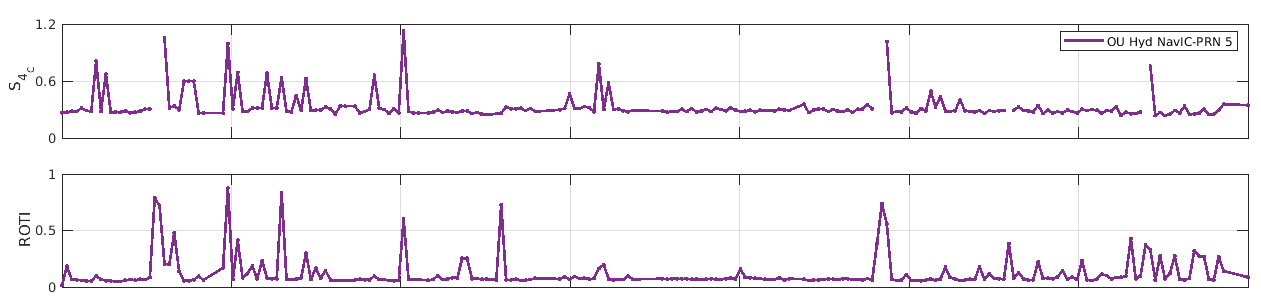}
\includegraphics[width=0.95\textwidth,height=2.2in]{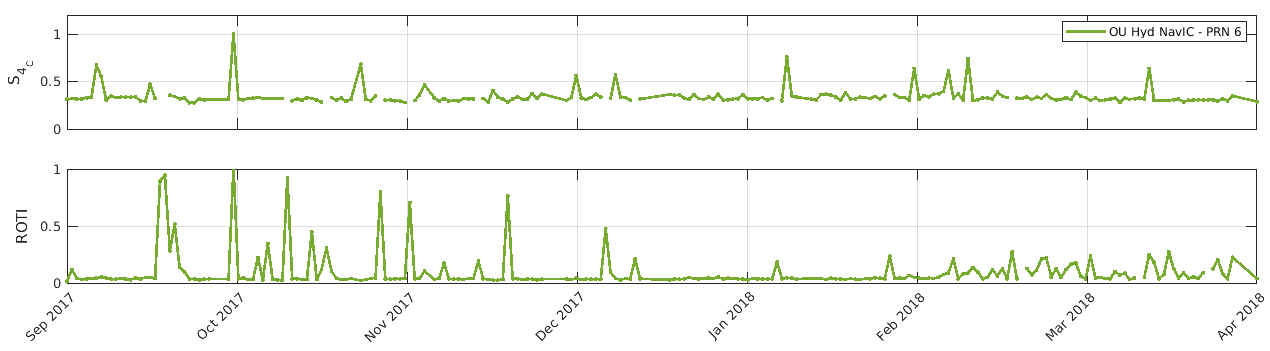}
\caption{The variation of F10.7 along with NavIC observed values of peak $S_{4_C}$, and ROTI by PRNs 2, 5 and 6 from Hyderabad station, for the analysis period: September 2017 through April 2018.}
\label{fig:Fig4}
\end{figure}

To extract scintillation events from the GPS data and based on evaluations of non scintillation baseline indicators as illustrated by \citep{41,40} in support with NOAA scales \citep{26}, 27 days are selected out of 607 days Fig: \ref{fig:Fig2}, \ref{fig:Fig3} and \ref{fig:Fig4}) analyzed spanning from September 01, 2017 to September 30, 2019 for the two stations, the following criteria are given below:

\begin{itemize}

    \item  The masking angle is set to $20^o$ to minimize multipath effects for the elevation in data sets for both NavIC and GPS.
    \item  The data of $S_{4_C}$ is filtered in such a way that the threshold value is 0.3.
    \item The event doesn't qualify as a scintillation event unless it remains above this threshold value for a period of at least 30 seconds, and any event which initiates within an elapsed time of five minutes after the previous event is not considered a separate event to avoid certain interference effects.
    \item The $S_{4_C}$ events observed from multiple satellites simultaneously are analyzed as separate events.
    \item The events for which the loss-of-lock (LoL) has lasted for more than 120 seconds are evaluated for the analysis.
\end{itemize}
In this work, the LoL is defined as the ability of the receiver to lose track of a transmitting satellite. During this time interval, the receiver will be unable to log any phase or pseudorange observation for that duration.
Following the above set of criteria and analyzing the entire 25 months of observations of NavIC Indore, only 27 days qualify as scintillation events (Table \ref{tab1}). However, in the analysis, only 4\% of the nights in the total duration of the scintillations were detected and these events have been selected based on the distribution along the studied period.
The values of sunspot number (SSN), the F10.7 solar radio flux (s.f.u.), and the Dst index (nT), as well as peak values of $S_{4_C}$ and ROTI for the 27 days of filtered data based on the selection criteria, are shown in Table \ref{tab1}.

\begin{table}
\centering
\caption{The $S_{4_C}$ occurrences observed for the 27 days, based on the selection criteria of scintillation events, with the corresponding values of the sunspot number (SSN), the F10.7 solar radio flux (s.f.u.), the Dst index (nT), and the peak values of $S_{4_C}$ and ROTI over Indore and Hyderabad as observed by NavIC and GPS over Indore (IITI) and Hyderabad (OU).}

\vspace{10pt}
\resizebox{1.0\textwidth}{0.4\textheight}{
\begin{tabular}{|p{1.9cm}|p{0.8cm}|p{1cm}|p{0.8cm}|p{1.1cm}|p{1.2cm}|p{1.2cm}|p{1.1cm}|p{1.1cm}|p{1.1cm}|}
\hline
\textbf{Date (yy-mm-dd)} & \textbf{SSN} & \textbf{F10.7 s.f.u} & \textbf{Dst (nT)}& \textbf{peak $S_{4_C}$ GPS IITI} & \textbf{peak $S_{4_C}$ NavIC IITI} & \textbf{peak $S_{4_C}$ NavIC OU}&\textbf{peak ROTI GPS IITI} & \textbf{peak ROTI NavIC IITI} & \textbf{peak ROTI NavIC OU}\\
 \hline
17-09-08& 88& 188.5& -124& 0.65& 0.61& 0.80& 0.07& 0.10& 0.1\\
17-09-16& 13& 72.9&	  -30& 0.30& 0.35& 0.29& 0.10& 0.81& 0.06\\ 
17-09-27& 37& 91.3&	  -46& 1.10& 2.57&    -& 0.11& 0.55& -\\
17-10-12&  0& 69.9&	  -41& 0.96& 0.63& 0.59& 0.09& 0.70& 0.34\\
17-10-24& 23& 76.7&	  -28& 0.33& 1.20& 0.33& 0.09& 0.32& 0.30\\
17-10-25& 23& 77.9&	  -24& 0.30& 0.33&  0.33& 0.03& 0.18& 0.14\\
17-12-01& 23& 68.3&	   -8& 0.32& 0.34&  0.32& 0.08& 0.07& 0.064\\
17-12-04& 23& 66.4&	  -45& 0.33& 0.48&     -&  0.1& 0.77& -\\
17-12-14&  0& 69.8&	   -7& 0.75& 0.64& 0.56& 0.09& 0.6& 0.064\\
17-12-27& 13& 68.6&	  -11& 0.33& 1.39& 0.31& 0.09& 0.19& 0.06\\
18-02-16& 12& 69.8&	  -16& 0.36& 1.29&    -& 0.13& 0.91& -\\
18-02-19&  0& 67.5&	  -20& 0.33& 0.43& 0.33& 0.09& 0.36& 0.05\\
18-02-20&  0& 66.3&	  -28& 0.35& 0.98& 0.30& 0.1& 0.25& 0.06\\
18-02-21&  0& 68.7&	  -15& 0.33& 0.40& 0.30& 0.12& 0.30& 0.11\\
18-02-22&  0& 67&	  -13& 0.30& 0.31& 0.31& 0.10& 0.13& 0.05\\
18-02-24&  0& 66.8&	  -29& 0.35& 0.34& 0.28& 0.10& 0.45& 0.05\\
18-02-25&  0& 65.9&   -15& 0.35& 0.45& 0.36& 0.10& 0.15& 0.06\\
18-02-27& 15& 66.6&   -30& 0.38& 0.43& 0.30& 0.09& 0.65& 0.07\\
18-02-28& 12& 67.5&	  -20& 0.40& 0.38& 0.37& 0.11& 0.09& 0.12\\
18-03-01& 12&   66&   -17& 0.42& 0.31&    -& 0.11& 0.65& -\\
18-03-02& 11& 66.6&	  -12& 0.45& 0.35&    -&  0.1& 0.12& -\\
18-03-13& 0&  67.7&	   -5& 0.34& 1.01&    -& 0.12& 0.24& -\\
18-03-22& 0&	68&	  -28& 0.41& 1.23& 2.01& 0.11& 0.24& 0.12\\
18-04-28& 0&  71.1&	   -1& 0.34& 0.36&    -& 0.10& 1.13& -\\
18-05-28& 20& 78.9&	   13& 0.37& 0.43&    -& 0.1& 0.12&  -\\
18-06-05&  0& 73.4&   -15&  0.39& 0.34&   -& 0.11& 0.18& -\\
18-09-26&  0&	69&	  -14&  0.37& 0.42&   -& 0.10& 0.55& -\\
\hline
\end{tabular}}
\label{tab1}
\end{table}

\clearpage

\clearpage

Fig. \ref{fig:Fig2}, \ref{fig:Fig3} and \ref{fig:Fig4} represent the overall variation of F10.7 flux along with peak $S_{4_C}$ and ROTI for GPS Space Vehicle Identification (SVID) 1 over Indore and the NavIC PRNs 2, 5 and 6 over Indore and Hyderabad respectively. The blank spaces in these plots indicate data gaps during that period. It is found that the intensity of scintillation, as indicated by $S_{4_C}$, gradually increased during the equinoctial months (March–April, September–October). The scintillation events were rarely observed in the other months (where the F10.7 flux values were low) of the period of analysis in both stations. In support of the former observation, the corresponding values of ROTI, which are also known as a proxy for the $S_{4_C}$ index, also show the same variation. During the equinoctial months, the ROTI values for all the satellites of both the constellations varied between 0.5 to 1 for most of the days. The seasonal and solar activity dependence of the scintillation occurrence as observed using NavIC over these regions is in agreement with that of previous studies \citep{55,56,57,58}.

\subsection{Analysis of a typical scintillation event}

\begin{figure}
\centering
\includegraphics[width=\textwidth,height=5.5in]{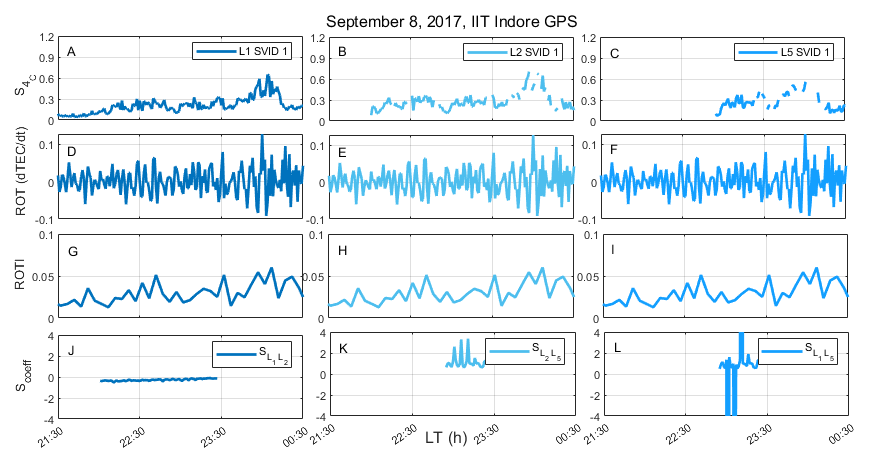}
\caption{The $S_{4_C}$, ROT, ROTI, and $S_{coeff}$ variations as observed from Indore station during during 21:30 LT (h) of September 08, 2017 to 00:30 LT (h) of September 09, 2017 for GPS SVID 1 for all three L-band frequencies.}
\label{fig:Fig5}
\end{figure}

\begin{figure}
\centering
\includegraphics[width=\textwidth,height=5.5in]{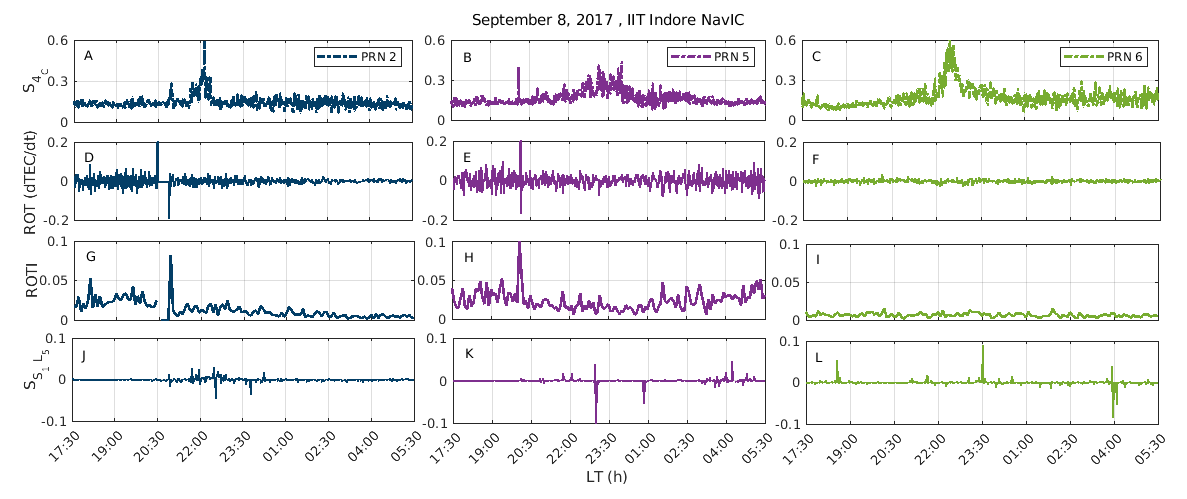}
\caption{The $S_{4_C}$, ROT, ROTI and $S_{S_1}{L_5}$ variation on September 08, 2017 as observed from Indore station during 17:30 LT (h) of September 08, 2017 to 05:30 LT (h) of September 09, 2017 for NavIC PRNs 2,5 and 6.}
\label{fig:Fig6}
\end{figure}

\begin{figure}
\centering
\includegraphics[width=\textwidth,height=5.5in]{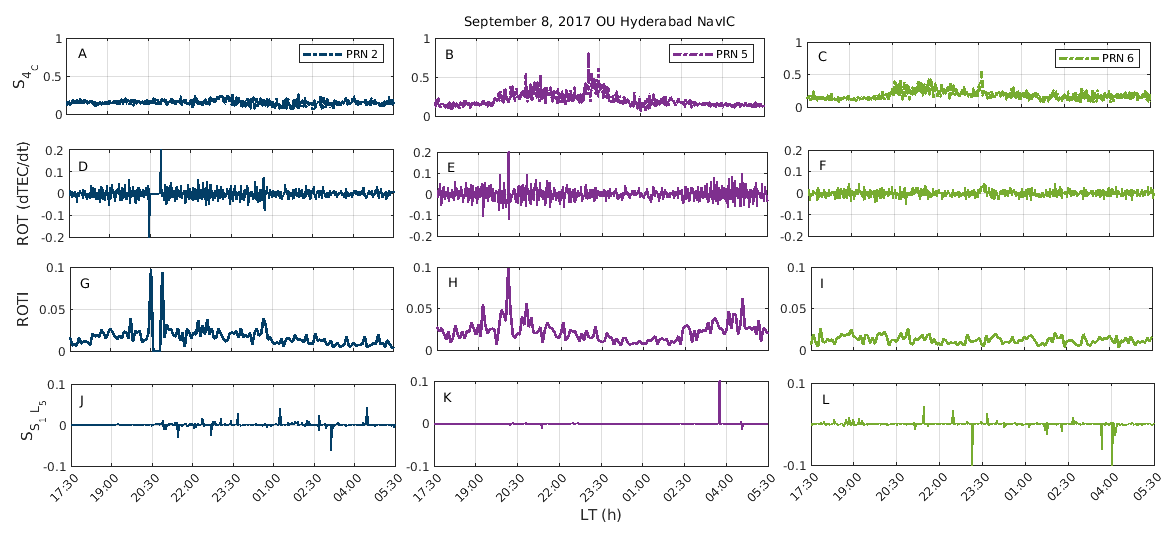}
\caption{The $S_{4_C}$, ROT, ROTI and $S_{S_1}{L_5}$ variation on September 08, 2017 as observed from Hyderabad station during 17:30 LT (h) of September 08, 2017 to 05:30 LT (h) of September 09, 2017 for NavIC PRNs 2,5 and 6.}
\label{fig:Fig7}
\end{figure}

This subsection describes a typical case of scintillation event observed on September 08, 2017, based on the formulae provided to convert the $C/N_0$ to scintillation index $S_{4_C}$ (equations \ref{appeneqnc1} to \ref{appeneqnc4} ) \citep{20,21,22}.
Furthermore, the rate of change of slant TEC (ROT), introduced by \citep{43}, provides an estimate of the ionospheric disturbances, by taking into account the average standard deviation of the rate of slant TEC. This index is defined as the Rate of TEC Index (ROTI). The formulae used to calculate values of ROT and ROTI are presented in Appendix (equations \ref{appeneqnd1} and \ref{appeneqnd2} respectively). The $S_{4_C}$, the ROT (dTEC/dt), the ROTI, and the scattering coefficients \citep{42} (see \ref{appeneqne1}) are plotted as a function of LT (h) in Fig. \ref{fig:Fig5}, \ref{fig:Fig6} and \ref{fig:Fig7} for GPS and NavIC observations over Indore and NavIC observations over Hyderabad respectively, on September 08, 2017.

The dominant mechanism, in the onset of amplitude scintillations, is the Pre-reversal Enhancement generated as a result of the enhancement in the vertical E$\times$B drift due to the eastward electric field at the sunset terminator (where the terminator implies a moving boundary between the day side and night side regions on the Earth) generating EPBs and ESFs around this time \citep{63,64,66}. \citep{62} in their study have brought forward the relationship between the onset of low-latitude scintillation and the sunset terminator. It is also to be noted that the onset of scintillation at a given location represents the time when the irregularities, that produce these scintillations, are initially generated \citep{65}. Fig. \ref{fig:Fig5} (A to C) shows the onset times of amplitude scintillation $S_{4_C}$ as observed from GPS SVID 1 at $L_1$, $L_2$, and $L_5$ frequencies during the local time between 23:30 and 00:30 LT (h). Fig. \ref{fig:Fig5} (D to F) and (G to I) show the ROT and ROTI values as observed from the same SVID during the occurrence time of $S_{4_C}$. Fig. \ref{fig:Fig5} (J to L) depicts the variation of scattering coefficients ($S_{coeff}$) from all three frequencies based on the deviations estimated from each of the frequencies of GPS. A clear deviation from the mean value is observed, supporting the fact that signal strength is affected during $S_{4_C}$ events, especially for the GPS frequency combination ($S_{coeff}$) of GPS ${L_2L_5}$ and GPS ${L_1L_5}$. Similarly, Fig. \ref{fig:Fig6} and \ref{fig:Fig7} show the onset times as observed from NavIC PRNs 2, 5, and 6 over Indore and Hyderabad, respectively. The onset times were observed to be around 20:30 LT (h) and 19:30 LT (h) respectively for Indore and Hyderabad, while the corresponding peak values of $S_{4_C}$ at these locations vary between 22:00 and 23:30 LT (h) for the PRNs 2, 5 and 6 of NavIC. It is clear from Fig. \ref{fig:Fig5},\ref{fig:Fig6} and \ref{fig:Fig7} that the $S_{4_C}$ oscillations match with the oscillations observed in ROT, ROTI and $S_{coeff}$. Although in Fig. \ref{fig:Fig6} and \ref{fig:Fig7} NavIC PRN 6 has no clear indicators of ROT and ROTI for both the stations Indore and Hyderabad it is still clear that $S_{4_C}$ oscillations match with the deviations of $S_{coeff}$.
In addition to these observations, it is to be noted that values of $S_{4_C}$ spiked up on these days, particularly for the PRNs 2, 5, and 6. This trend is noted during the whole period of the analysis presented in this work for the two stations.
Additionally, the oscillating values observed in the $S_{S_1 L_5}$ match with the fluctuations of $S_{4_C}$. The work presented here clearly shows that NavIC amplitude fluctuations present a pattern in the onset time of scintillation detection. During the period of analysis, for all the days, the difference in the onset time of scintillation was mostly during the post-sunset period when compared to pre-sunrise events. However, as shown by \citep{55}, a similar pattern in the scintillation onset time has been observed for the GPS signals over the Brazilian sector, which is near to southern crest of EIA, with an average onset period of 30 to 60 minutes before the onset time observed in this work. The peak values and onset time of scintillations at various stations covering the Indian region using GPS and Digi-sonde data, reported by \citep{56,59} showed that the scintillation onset time occurred between 19:30 LT and 23:30 LT, closely matching the current results. In support of the present observations, the variation of the onset time of scintillation as presented in \citep{57,58} showed that measurements of scintillation using GPS satellites cover a wide longitude region, and the onset time difference of scintillation during a specified season is linked with the different locations where scintillation was initially generated, especially near the northern crest of EIA.

\clearpage

\clearpage

\clearpage

\newpage
\section{Characterizing Amplitude Scintillation}

After identifying the scintillation events over this period of 25 months, we proceeded towards characterizing the scintillation events observed from NavIC Indore and validated by NavIC Hyderabad and GPS Indore. To characterize amplitude scintillations at low-latitudes, earlier studies \citep{44,45} found evidence to evaluate statistical models. The Nakagami-m distribution was found to be the optimum fit for the data set in the study results. A decade later according to \citep{46}, theoretical Nakagami-m distribution was found to be the best for explaining different levels of fade. In the early 1990s, this was one of the first studies in the Indian region, with a C and L band transionospheric scintillation. \citep{47} proposed a simple model for simulating equatorial transionospheric radio wave scintillation, and this study concluded that the Rice distribution was considered to be the best fit for the experimental data. Following these experiments, \citep{48,49,50} identified the benefits of using a two-parameter model, recommending the use of the $\alpha-\mu$ and Nakagami-m distributions. The studies proposed a model based on the $\alpha-\mu$ distribution parameters and taking into account the amplitude and phase scintillation correlation \citep{51}. The numerous findings presented in this study clearly illustrate how, under the influence of ionospheric conditions, a receiver experiences different error values leading to a specified scintillation level. $S_{4_C}$. \citep{51} introduced a $\alpha-\mu$ distribution which in general is a fading model, which explores the non-linearity of the propagation medium and is a revised form of the Stacy (generalized Gamma) distribution. \citep{48} proposed the use of this distribution for modeling ionospheric amplitude scintillation.

The model is described by the coefficients $\alpha$ and $\mu$ of the normalized amplitude envelope of the received signal ($r$), where $\alpha$ is the modulus of the sum of the multipath components and $\mu$ is the number of multipath components and the $\alpha-\mu$ probability density function, assuming that the average signal power (or intensity) $r^2$ is equal to 1 for $E[R^2] = 1$ is defined as:
\begin{equation}
f(r)=\frac{\alpha{r^{\alpha\mu-1}}}{{\zeta^{\alpha\mu/2}}{\Gamma(\mu)}} exp \big(-\frac{r^{\alpha}}{\zeta^{\alpha/2}}),     
\label{Eqn:Eqn1}
\end{equation}

where in
\begin{equation}
     \zeta = \frac{\Gamma(\mu)}{\Gamma (\mu + 2/\alpha)}
\end{equation}
$\Gamma(.)$ is the gamma function. \citep{51} used this model to describe the mobile communication channel, assuming that the received signal is the consequence of a cluster of multipath waves propagating in an heterogeneous environment. The most interesting feature of this model is that it can simulate a number of different scenarios depending on the values of its variables, for instance, the $\alpha-\mu$ distribution can be transformed into a Nakagami-m distribution ($\alpha$ = 2 and $\mu$ = m, where m is the single parameter of the distribution), a Rayleigh distribution ($\alpha$ = 2 and $\mu$ = 1), or a Weibull distribution (with two parameters $\alpha$ and $\gamma$ =$\zeta^{1/2}$, for $\mu$ = 1).

As described in Appendix, $S_4$ index characterizes the strength of amplitude scintillation where the intensity of received signal $I=|r^2|$. The Nakagami-m parameter in the Nakagami distribution can be related to $S_4$ index by the following notation where $m = {1}/{S_4^{2}}$ and the relation to estimate this is given below
\begin{equation}
m =  \frac{{E}^2({r}^2)}{{{E}({r}^4)} - {E}^2({r}^2)}
\label{Eqn:Eqn2}
\end{equation}

The $\alpha$ and $\mu$ coefficients can be calculated based on an experimental study conducted by \citep{51}.

\begin{equation}
\frac{{E}^2({r}^\beta)}{{{E}({r}^{2\beta})} - {E}^2({r}^\beta)} = \frac{\Gamma^{2} (\mu + \beta/\alpha)}{\Gamma(\mu)\Gamma (\mu + 2\beta/\alpha)-\Gamma^{2} (\mu + \beta/\alpha)}
\label{Eqn:Eqn3}
\end{equation}

The left-hand side of the equation (\ref{Eqn:Eqn3}) can be derived from field data for randomly chosen values of the parameter $\beta$, which provides the order of the system of $r$ to be determined. In fact, when the left-hand side of equation (\ref{Eqn:Eqn3}) for $\beta$ = 2 is compared to the right-hand side of equation (\ref{appeneqnc2}), the following results are obtained:

\begin{equation}
{S_4}^2 = \frac{\Gamma(\mu)\Gamma (\mu + 4/\alpha)-\Gamma^{2} (\mu + 2/\alpha)}{\Gamma^{2} (\mu + 2/\alpha)}
\label{Eqn:Eqn4}
\end{equation}
If $\alpha$ = 2 in the above equation (\ref{Eqn:Eqn4}) and the properties of the gamma function are used, the value of the left-hand side of the same equation implies $1/\mu = 1/m$, which is the condition for the Nakagami-m distribution (\citep{48}).

\begin{figure}
\centering
\includegraphics[width=\textwidth,height=3.5in]{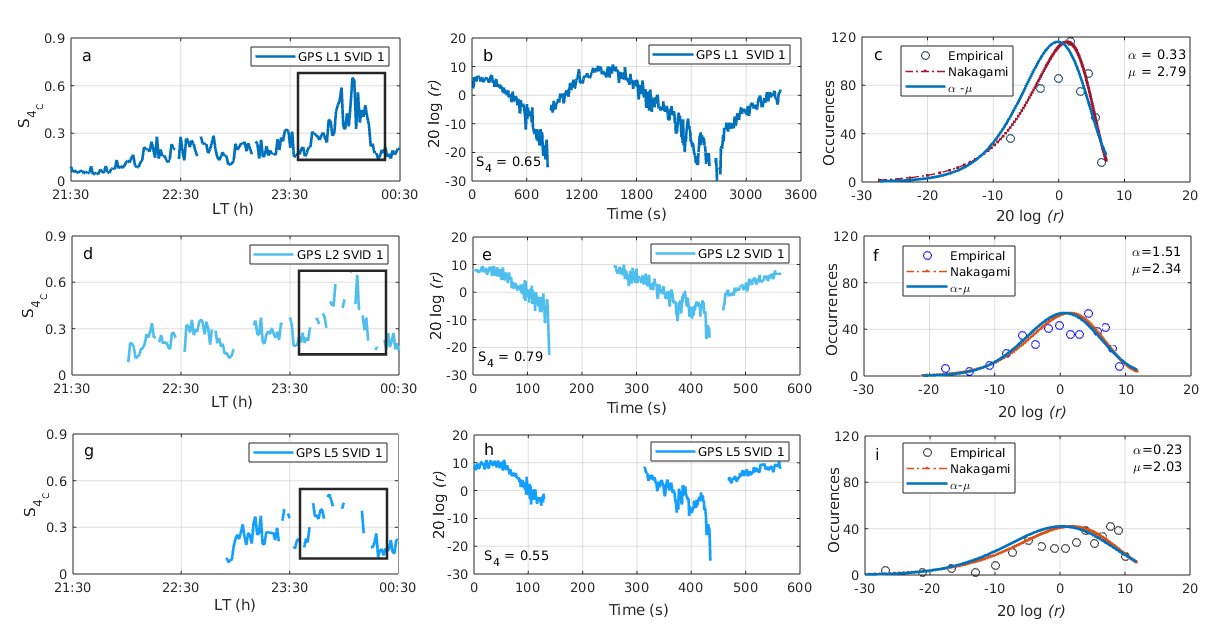}
\caption{Scintillation event on September 08, 2017, as observed by GPS SVID 1 $L_1$, $L_2$, and $L_5$ signals over Indore. Panels a, d, and g show variations of $S_{4_C}$ for a portion of the day. The black rectangular box denotes the epochs when scintillation was detected. Panels b, e, and h shows fading signal intensity for the period marked with a rectangular box in panels a, d and g. Panels c, f and i show the calculated occurrences along with the $\alpha-\mu$ distributions and the Nakagami-m distribution (refer to equations 1 to 5)}
\label{fig:Fig8}
\end{figure}

Applying the same to characterize amplitude scintillation, analysis for a chosen event of scintillation on September 08, 2017, where the $C/N_0$ value has significantly dropped below 30 dB-Hz and has approached zero between 22:00 to 0:00 LT (h) as observed from both the constellations, is considered. During the event, only SVID 1 from the GPS constellation was able to capture the intense value of amplitude scintillation. The corresponding signal intensity fading (20log($r$)) for Indore GPS SVID 1 during the $C/N_0$ drop period, lasted for the period of 3400, 600, and 600 seconds for $L_1$, $L_2$, and $L_5$, respectively. The $S_{4_C}$ index as a function of local time is shown for each of the frequencies in Fig. \ref{fig:Fig8} (a, d and g) respectively.

\begin{figure}
\centering
\includegraphics[width=\textwidth,height=3.5in]{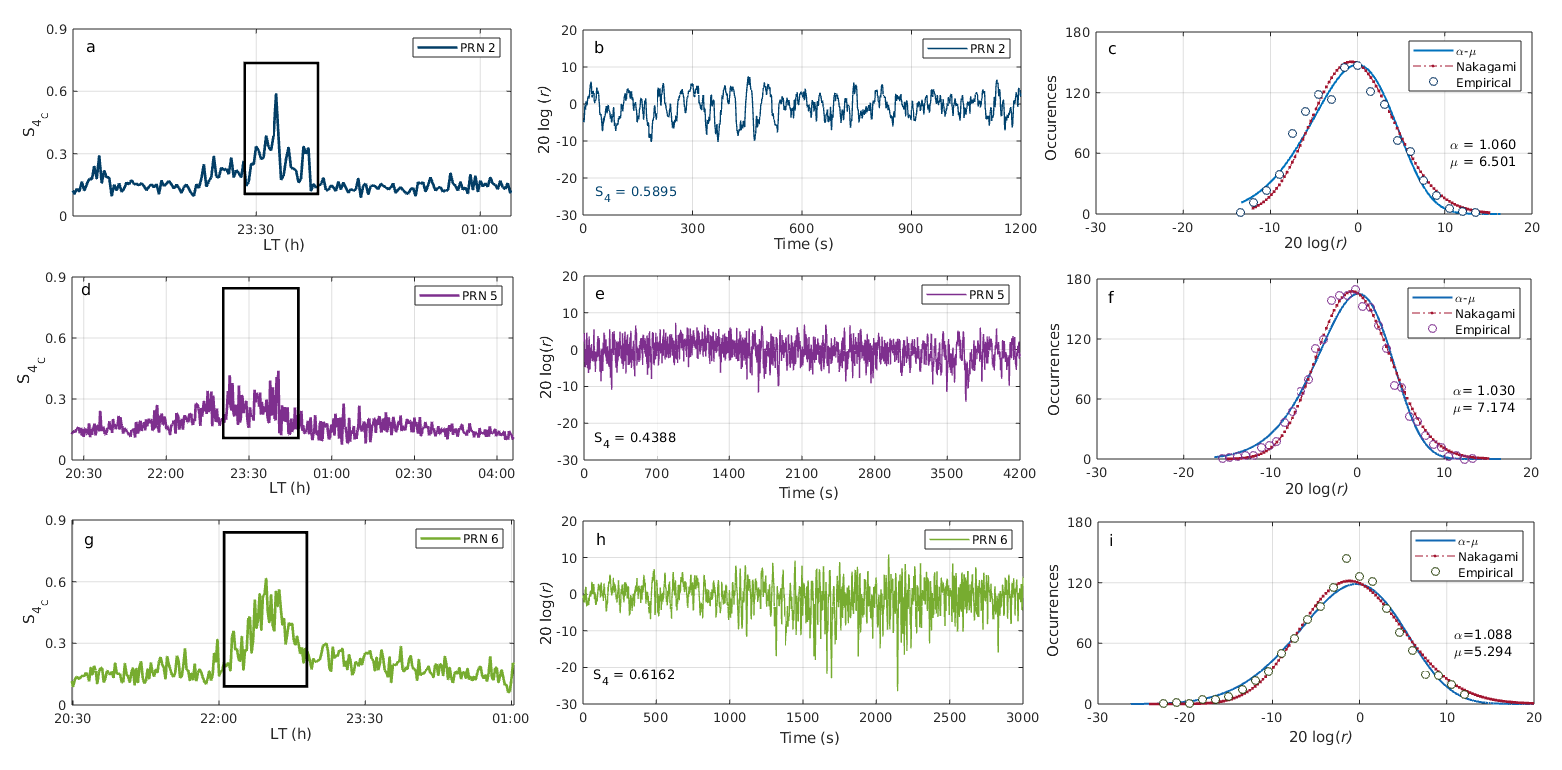}
\caption{Scintillation event on September 08, 2017, as observed by NavIC PRNs 2,5 and 6 signals over Indore. Panels a, d, and g show variations of $S_{4_C}$ for a portion of the day. The black rectangular box denotes the epochs when scintillation was detected. Panels b, e, and h shows fading signal intensity for the period marked with a rectangular box in panels a, d and g. Panels c, f and i show the calculated occurrences along with the $\alpha-\mu$ distributions and the Nakagami-m distribution (refer to equations 1 to 5)}
\label{fig:Fig9}
\end{figure}

\begin{figure}
\centering
\includegraphics[width=\textwidth,height=3.5in]{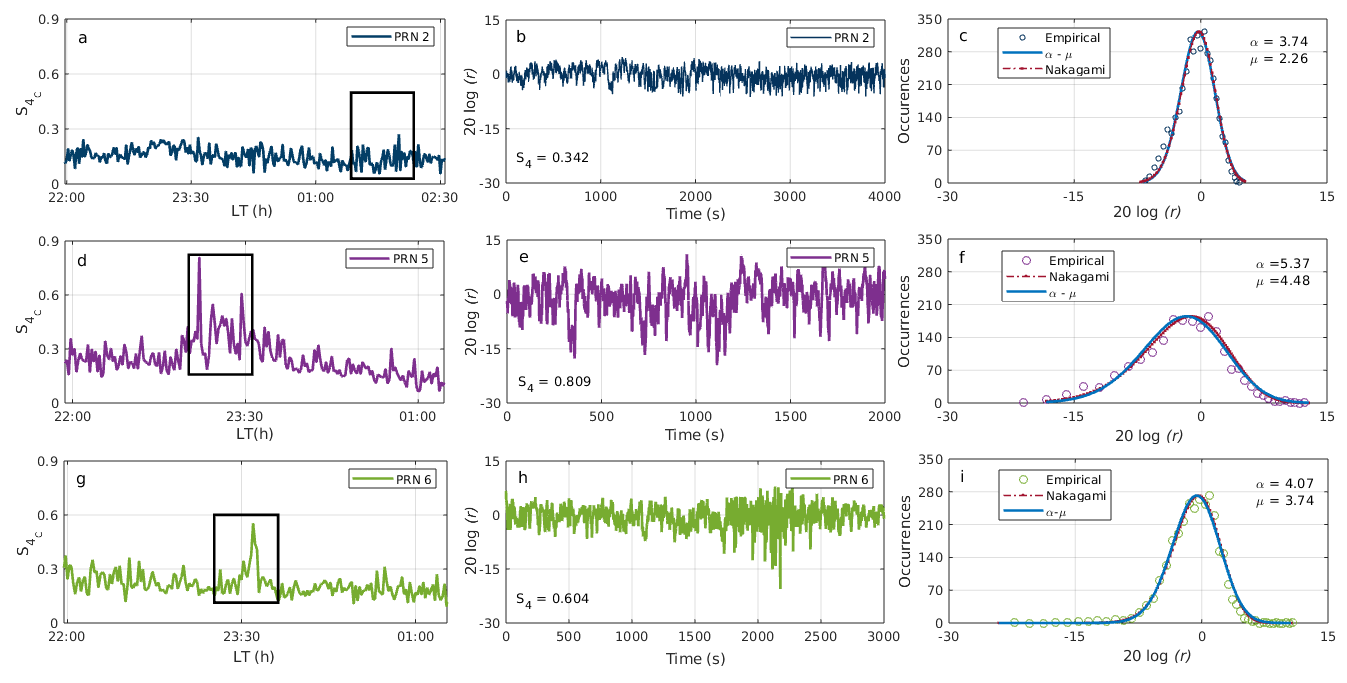}
\caption{Scintillation event on September 08, 2017, as observed by NavIC PRNs 2, 5, and 6 signals over Hyderabad. Panels a, d, and g show variations of $S_{4_C}$ for a portion of the day. The black rectangular box denotes the epochs when scintillation was detected. Panels b, e, and h shows fading signal intensity for the period marked with a rectangular box in panels a, d and g. Panels c, f and i show the calculated occurrences along with the $\alpha-\mu$ distributions and the Nakagami-m distribution (refer to equations 1 to 5)}
\label{fig:Fig10}
\end{figure}

The values where the amplitude scintillation peaks are marked (in a rectangular box) to indicate the fading sample that is used to calculate the signal intensity fading (20log($r$)) as a function of time in seconds in Fig. \ref{fig:Fig8} (b, e, and h). The corresponding values of $\alpha-\mu$ and Nakagami distribution for the scintillation occurrences as a function of 20 log ($r$) are shown in Fig. \ref{fig:Fig8} (c, f and i) respectively for $L_1$, $L_2$, and $L_5$ frequencies. The scintillation occurrences are the empirical data in the plots for which the distribution curves have been plotted in Figs. \ref{fig:Fig8},\ref{fig:Fig9} and \ref{fig:Fig10}. In the case of GPS-SVID 1, though the rectangular box appears to be for a longer time interval, the instances of discontinuous data at $L_2$ and $L_5$ frequencies reduce the fading sample to be less than $L_1$. The NavIC (Indore and Hyderabad) observations on the same day for the $S_{4_C}$ indices for PRN 2, 5 and 6 are shown in Fig. \ref{fig:Fig9} and \ref{fig:Fig10} (a, d and g) along with the sample of data points chosen to calculate corresponding signal fading for these PRNs in Fig. \ref{fig:Fig9}, \ref{fig:Fig10} (b, e and i). The distribution curves for $\alpha-\mu$ and Nakagami for the occurrences (empirical data) here in the case of NavIC are plotted in Fig. \ref{fig:Fig9}, \ref{fig:Fig10} (c, f and i) respectively. The sample windows in the above observations and corresponding Fig. \ref{fig:Fig8},  \ref{fig:Fig9} and \ref{fig:Fig10} show how the severity of fading occurrences increases as the value of $S_{4_C}$ rises. The results for the values of $\alpha-\mu$ parameters also show that $\alpha$ increases with the increase in $S_{4_C}$. This depicts how challenging it is for the receiver to retain the lock as the signal quality degrades. As a result, defining the statistics of these fades is important for estimating the effects of these phenomena and supporting the development of signal processing techniques and positioning algorithms that will reduce these effects in the receiver. The observations presented will be used as the starting points for further studies related to our evaluation of positional accuracy with the aid of NavIC under disturbed as well as quiet-time ionospheric conditions.
 
\newpage 
\section{Discussions and Conclusion} 

A systematic study of the variable low-latitude region around the EIA and the magnetic equator is always essential to probe the ionosphere as well as its impact on other aspects of communication and navigation. Here, a study in low solar activity conditions using NavIC is presented. It should be noted that in a previous study, see \citep{38}, it has been shown that NavIC has significant advantages over the GPS for ionospheric studies in the Indian subcontinent region. These observations, when combined with the GPS constellation, and the additional ray path through the ionosphere provided by this system, help to better observe space weather effects. The observations presented show NavIC's efficiency, in studying space weather, which has been evaluated at Indore using 27 critical days out of 607 days of observation. Detailed case studies, from two different stations (Indore: located near the northern crest of EIA, and Hyderabad: located near the magnetic equator), have been performed to validate the reliability of NavIC to study the ionosphere during space weather events. It has been found that the variation of $S_{4_C}$ registered from the GPS closely matches that of NavIC $S_{4_C}$. It is even more interesting to observe the post-sunset ionospheric response of NavIC PRNs 2, 5, and 6 w.r.t the $S_{4_C}$ index. The values of NavIC $S_{4_C}$ and the ROT, as well as ROTI indices, have a clear signature of the occurrence of ionospheric scintillation. The scintillation onset times have been observed to be occurring around 19:30 LT (h) and 20:30 LT (h) for Hyderabad and Indore, respectively, and the peak values of $S_{4_C}$ to be occurring between 22:00 and 23:00 LT (h). The $S_{4_C}$ proxy parameters: ROT, peaked at 0.2 and ROTI, peaked at 0.1. The fluctuations in Fig. \ref{fig:Fig3} are more when compared to the ones in Fig. \ref{fig:Fig4} and the reason behind the variation in fluctuations is still not known and can be answered well only when such an analysis is carried at various locations with NavIC and GPS. The results obtained in our study are in complete agreement with similar works presented in previous literature by \citep{2,8,9,11,13,23}. One of the important outcomes of our study is that it shows, for the first time, the existence of a strong correlation between NavIC signals from $L_5$ and $S_1$ (see Fig. \ref{fig:Fig5}, \ref{fig:Fig6}, and \ref{fig:Fig7}). We have shown the same using the scattering coefficients analysis where it is found that the scattering is minimum between the $L_5$ and $S_1$ signals of NavIC. This is in contrast with the previous study done by \citep{42} where they used $L_1$ and $L_5$ signals of GPS. Furthermore, studies presented here using empirical scintillation data show that the amplitude distribution is well modeled as Nakagami-m and $\alpha-\mu$ distributions as presented in \citep{45,46,47,48,49,50}. These models have been validated in comparison tests with empirical scintillation data for various PRNs of NavIC. A follow-up work, from this study based on a detailed characterization of the scintillation events during these 25 months using the $C/N_0$ observations from both NavIC and GPS, has also been performed. While the work presented here is limited to two NavIC stations, similar comparative studies with NavIC at different stations, under variable ionospheric conditions, will be carried out in the future to validate our results over large geographic locations of the Indian subcontinent.

\newpage
\section*{Acknowledgments} 

DA acknowledges the INSPIRE fellowship from the Department of Science and Technology, which she used to pursue her research. SC acknowledges ISRO's Space Applications Center (SAC) for offering an NGP-17 research fellowship during his Ph.D. and the Department of Space for providing a fellowship during his post-doctoral research. The authors also thank SAC, ISRO for providing the NavIC ACCORD receiver to the Department of Astronomy, Astrophysics, and Space Engineering, IIT Indore, under the NGP-17 project. The authors wholeheartedly thank Dr. P. Naveen Kumar for providing NavIC Osmania University (OU) Hyderabad station data. DA is extremely thankful to Harsha Avinash Tanti and Anshuman Tripathi for the fruitful discussions and moral support throughout the analysis period. 

\newpage
\bibliographystyle{model2-names.bst}
\biboptions{authoryear}
\bibliography{S4dpt.bib}

\newpage
\appendix

\section{Calculations of $S_{4_C}$ index, the ROT, the ROTI, and the Scattering Coefficients} 

The study presented in this paper utilizes Carrier to Noise density ($C/N_0$) (dB-Hz) measurements of NavIC satellite signals to estimate signal intensity ($S_I$) as its primary data for the $S_{4_C}$ index values.
\begin{equation}
S_I= 10^{0.1*C/N_{0}}
\label{appeneqnc1}
\end{equation}
where the value of $S_I$ is denoted as the intensity of the signal. The $S_4$ index is then calculated  defined as the signal's normalised variance of intensity, which is expressed as:
\begin{equation}
{S_4} = \sqrt \frac{\langle{S_I}^2\rangle - {\langle S_I \rangle}^2} {{\langle S_I \rangle}^2}
\label{appeneqnc2}
\end{equation}
where \textit{$S_I$} is the intensity of the signal.
Thus calculated $S 4$ index still requires a correction factor as the $C/N_0$ value of any radio signal is inclusive of ambient noise and in order to calculate the $S_4$ by applying correction as below gives us ambient noise free $S_4$ index and is denoted as $S_{AN}$ index
\begin{equation}
S_{AN}= \sqrt\frac{100}{C/N_0} \Big[1+\frac{500}{19 C/N_0}\Big]
\label{appeneqnc3}
\end{equation}

Now subtracting equation (\ref{appeneqnc3}) from equation (\ref{appeneqnc2}) and the real estimate for $S_{4_C}$ is obtained as shown below
\begin{equation}
S_{4_C}= S_{4} - S_{AN}
\label{appeneqnc4}
\end{equation}

The ROT and ROTI values \citep{43} are calculated as given in equations below where in equation (\ref{appeneqnd1}) $ STEC_{r+1} $ and $ STEC_{r} $ are slant TEC at {\textit{r+1}} and {\textit{r}} time epochs; $\Delta t_{r}$  time interval; usually the unit of ROT is TECU/min and in equation (\ref{appeneqnd2}) where $\langle ROT \rangle$ denotes averaging ROT during {\textit{N}} epochs.
 \begin{equation}
{ROT} =  \frac{{STEC_{r+1}}-{STEC_{r}}} {\Delta t _ r} .
\label{appeneqnd1}
\end{equation}

\begin{equation}
{ROTI} = \sqrt {\langle{ROT}^2\rangle - {\langle ROT \rangle}^2}
\label{appeneqnd2}
\end{equation}

A scattering coefficient \citep{42} across a pair of frequencies was later defined as the difference of $C/N_0$ fluctuations normalized with respect to the amount of those fluctuations, based on these computed $C/N_0$ deviations. The formula for scattering coefficient is given by \ref{appeneqne1} where \textit{a} is the $C/N_0$ deviations calculated from one frequency is substituted and \textit{b} is the $C/N_0$ deviations calculated from second frequency respectively. Here $S_{a,b}$ is a dimensionless quantity with values near zero, suggesting a good similarity between the signals. Here, the $C/N_0$ deviations were determined by subtracting the moving averaged values over a ninety-minute running time period from the instantaneous $C/N_0$ calculation.
\begin{equation}
S_{a,b}  = \Big[\frac{a-b}{a+b}\Big]
\label{appeneqne1}
\end{equation}

The $C/N_0$ deviations of NavIC $S_1$ and $L_5$ signals are calculated. The scattering coefficients as observed from NavIC signals $L_5$ and $S_1$  show a very strong relationship during the quiet period of the ionosphere and even during the scintillation period the scattering coefficients values have not oscillated beyond the value of 0.1 for both observations over Indore and Hyderabad Fig. \ref{fig:Fig6} and Fig. \ref{fig:Fig7} of Section 4.

\end{document}